

\input harvmac
\def\ie{{\it i.e.}}

\def\no{\noindent}
\def\o{\over}

\def\nl{\hfill\break}

\def\ZZ{{\bf Z}}

\def\semidir{\times}
\def\ontopss#1#2#3#4{\raise#4ex \hbox{#1}\mkern-#3mu {#2}}

\setbox\strutbox=\hbox{\vrule height12pt depth5pt width0pt}
\def\tablerule{\noalign{\hrule}}
\def\tr{\tablerule}
\def\strut{\relax\ifmmode\copy\strutbox\else\unhcopy\strutbox\fi}

\nref\rKedMc{R.~Kedem and B.M.~McCoy, Stony Brook preprint ITP-SB-92-56,
  hep-th/9210129.}
\nref\rFLa{V.A.~Fateev and S.L.~Lykyanov, Int.~J.~Mod.~Phys.~A3
 (1988) 507.}
\nref\rJMO{M.~Jimbo, T.~Miwa and M.~Okado, Nucl.~Phys.~B300[FS22]
 (1988) 74.}
\nref\rCR{P.~Christe and F.~Ravanini, Int.~J.~Mod.~Phys.~A4 (1989)
 897.}
\nref\rFLb{V.A.~Fateev and S.L.~Lykyanov, Sov.~Sci.~Rev.~A Phys.~15
 (1990) 1.}
\nref\rAlt{D. Altschuler, M. Bauer and H. Saleur, J. Phys.A 23 (1990) L789.}
\nref\rKaPe{V.G.~Kac and D.H.~Peterson, Adv.~in Math.~53 (1984) 125.}
\nref\rDQ{J.~Distler and Z.~Qiu, Nucl.~Phys.~B336 (1990) 533.}
\nref\rGord{B.~Gordon, Amer.~J.~Math.~83 (1961) 393;\nl
  G.E.~Andrews, Proc.~Nat.~Acad.~Sci.~USA~71 (1974) 4082.}
\nref\rFNO{B.L.~Feigin, T.~Nakanishi and H.~Ooguri, Int.~J.~Mod.~Phys.~A7,
  Suppl.~1A (1992) 217.}
\nref\rWalg{F.A.~Bais, P.~Bouwknegt, ~K.~Schoutens and M.~Surridge,
 Nucl.~Phys.~B304 (1988) 348 and 371.}
\nref\rlw{J. Lepowsky and R. L. Wilson, Proc. Natl. Acad. Sci. USA, 78
(1981) 7254.}
\nref\rLepPrim{J.~Lepowsky and M.~Primc, {\it Structure of the
 standard modules for the affine Lie algebra $A_1^{(1)}$},
 Contemporary Mathematics, Vol.~46 (AMS, Providence, 1985).}
\nref\rZFparaf{A.B.~Zamolodchikov and V.A.~Fateev, Sov.~Phys.~JETP 62 (1985)
 215.}
\nref\rourii{T.R.~Klassen and E.~Melzer, Nucl.~Phys.~B338 (1990) 485.}
\nref\rMSDVVV{G.~Moore and N.~Seiberg, Commun.~Math.~Phys.~123~(1989)~177;\nl
 R.~Dijkgraaf, C.~Vafa, E.~Verlinde and H.~Verlinde,
 Commun.~Math.~Phys. 123 (1989) 485.}
\nref\rRoCa{A.~Rocha-Caridi, in: {\it Vertex Operators in Mathematics
 and Physics},
 ed. J.~Lepowsky {\it et al} (Springer, Berlin, 1985).}
\nref\rKaWa{V.G.~Kac and M.~Wakimoto, Adv.~in Math.~70 (1988) 156.}
\nref\rDV{ J.L.~Cardy, Nucl.~Phys.~B270 (1986) 186; \nl R.~Dijkgraaf
 and E.~Verlinde, Nucl.~Phys.~B (Proc.~Suppl.) 5B (1988) 87.}
\nref\rNRT{W.~Nahm, A.~Recknagel and M.~Terhoeven, Bonn preprint,
   hep-th/9211034.}
\nref\rRiSz{B.~Richmond and G.~Szekeres, J.~Austral.~Math.~Soc.~(Series~A)~31
             (1981) 362.}
\nref\rLewin{L. Lewin, {\it Dilogarithms and associated functions}
 (MacDonald, London, 1958).}
\nref\rKBR{A.N.~Kirillov, Zap.~Nauch.~Semin.~LOMI 164 (1987) 121
 (J.~Sov.~Math.~47 (1989) 2450);
 V.V.~Bazhanov and N.Yu.~Reshetikhin, J.~Phys.~A23 (1990) 1477.}
\nref\rsums{A.N.~Kirillov and N.Yu.~Reshetikhin, J.~Phys.~A20 (1987) 1587.}
\nref\rFG{P. Fendley and P. Ginsparg, Nucl. Phys. B324 (1989), 549.}
\nref\rPZ{P.A. Pearce and X.K. Zhao (in preparation).}
\nref\rKlumPear{A.~Kl\"umper and P.A.~Pearce, J.~Stat.~Phys.~64
 (1991) 13; Physica A 183 (1992) 304.}
\nref\rFatZam{V.A.~Fateev and A.B.~Zamolodchikov, Phys.~Lett.~92A
 (1982) 37.}
\nref\rDasKedMc{S.~Dasmahapatra, R.~Kedem and B.M.~McCoy (in preparation).}
\nref\rMTW{B.M.~McCoy, C.A.~Tracy and T.T.~Wu, Phys. Rev. Lett. 38
 (1977) 793.}
\nref\rTanig{A.B.~Zamolodchikov, Adv.~Stud.~Pure Math.~19 (1989) 641.}
\nref\rZams{A.B.~Zamolodchikov and Al.B.~Zamolodchikov, Nucl. Phys.
 B379 (1992) 602.}

\Title{\vbox{\baselineskip12pt\hbox{ITP-SB-92-64, RU-92-51}
\hbox{hep-th/9211102} } }
{\vbox{\centerline{Fermionic Quasi-Particle Representations} \nl
 \centerline{for Characters of~
  ${(G^{(1)})_1 \times (G^{(1)})_1 \o (G^{(1)})_2}$ }}}

\vskip 6mm
\centerline{R.~Kedem,\foot{Institute for Theoretical Physics,
 SUNY, Stony Brook,  NY 11794-3840}~
 T.R.~Klassen,\foot{Department of Physics and Astronomy, Rutgers
 University, Piscataway, NJ 08854-0849} ~B.M.~McCoy,$^1$~ and ~E.~Melzer$^1$}

\vskip 17mm

\centerline{\bf Abstract}
\vskip 3mm

We present fermionic quasi-particle sum representations for some of
the characters (or branching functions) of ~${(G^{(1)})_1 \times
(G^{(1)})_1 \o (G^{(1)})_2}$ ~for all simply-laced Lie algebras $G$.
For given $G$ the characters are written as the partition function of
a set of rank~$G$ types of massless quasi-particles in certain charge
sectors, with nontrivial lower bounds on the one-particle momenta.
We discuss the non-uniqueness of the representation for the identity
character of the critical Ising model, which arises in both the $A_1$
and $E_8$ cases.

\Date{\hfill 11/92}
\vfill\eject

\newsec{Introduction}
\ftno=0

Recently new sum representations for branching functions of
the algebra $A_3^{(1)}$, which appear in the (non-diagonal) modular
invariant partition function of the coset conformal field theory (CFT)
${(A_3^{(1)})_1 \times (A_3^{(1)})_1 \o (A_3^{(1)})_2}$, have been
obtained from an analysis of Bethe's equations for the
antiferromagnetic 3-state Potts chain~\rKedMc.

To explain this result consider the sum

\eqn\SofX{  S_G(q) ~\equiv~ \sum_{m_1=0}^\infty \ldots \sum_{m_n=0}^\infty ~
    {q^{{\bf m} C_G^{-1} {\bf m}^t} \o (q)_{m_1} \ldots (q)_{m_n} }
   ~\equiv~ \sum_{{\bf m=0}}^\infty ~
    {q^{{\bf m} C_G^{-1} {\bf m}^t} \o (q)_{m_1} \ldots (q)_{m_n} } ~~,}
where ${\bf m}=(m_1,\ldots,m_n)$, $(q)_m=\prod_{k=1}^{m}(1-q^k)$
for $m>0$, ~$(q)_0=1$, and $C_G$ is the Cartan matrix of the
simply-laced Lie algebra $G$ of rank $n$.
In ref.~\rKedMc\ it was found (and verified up to
$O(q^{200})$) that

\eqn\SofAiii{ \eqalign{S_{A_3}(q) ~&\equiv~
  \sum_{{\bf m=0}}^\infty {q^{(3m_1^2+4m_2^2+3m_3^2+4m_1 m_2
      +2m_1 m_3+4m_2 m_3)/4} \o (q)_{m_1} (q)_{m_2} (q)_{m_3} } \cr
  &=~ q^{1/24}~ [b^0_0(q;3)+2b^0_2(q;3)+b^0_4(q;3)] ~~,\cr} }
where the $b^l_m(q;n)$ are the branching functions of ${(A_n^{(1)})_1
\times (A_n^{(1)})_1\o(A_n^{(1)})_2}$~\rFLa\rJMO\rCR\rFLb, which are equal
by level-rank duality~\rAlt\ to the branching functions of
$(A_1^{(1)})_{n+1}/U(1)$ \rKaPe\rDQ\ given in (2.4) below.  Moreover,
restricting the sum on the first line of
\SofAiii\ to
\eqn\rest{m_1+2m_2+3m_3 \equiv Q~({\rm mod}~4),\qquad \hbox{with}\
    Q=0,\pm 1,2,}
 gives separately $b^0_0$, $b^0_2$, and $b^0_4$ (see eqs.~(3.13) and
(3.19) of~\rKedMc, where the notation ${\bf
m}=(m_{2s},m_{ns},m_{-2s})$ was used).
The sum representations of other characters
in this theory are also
presented in~\rKedMc.

Here we present the $ADE$ generalization of~\SofAiii\ and~\rest,
corresponding to the cosets ${(G^{(1)})_1 \times (G^{(1)})_1 \o
(G^{(1)})_2}$, where $G$ is an arbitrary simply-laced Lie algebra.
The possibly restricted sums of the form~\SofX\ again are branching
functions, and we discuss their interpretation.
 Similar expressions for characters of
the Virasoro minimal CFTs $M(2,2n+3)$ as the sum side of the Andrews-Gordon
identities~\rGord, which may be thought of as associated with the
algebras $A_{2n}^{(2)}$, have appeared in ref.~\rFNO.

\bigskip
\newsec{$ADE$ Generalization}

The (restricted) sums~\SofX\ will turn out to be characters of the
identity and certain other extended primary fields in the
corresponding coset CFTs, which are the first models in the ${\cal
W}G$ unitary series~\rFLb\rWalg.  In the $A_n$ case these CFTs are
also known to describe $\ZZ_{n+1}$
parafermions~\rlw\rLepPrim\rZFparaf\
of central
charge $c_n={2n \o n+3}$.  In the $D_n$ case they are~\rFLb\rourii\
the points $r=\sqrt{{n\o 2}}$ on the orbifold line (or gaussian line,
depending on the modular invariant combination of characters in the
partition function) of $c=1$ CFTs.  These CFTs are known to be minimal
with respect to extended chiral algebras~\rMSDVVV\ whose characters
are essentially theta functions, see below.  The CFTs in the $E_n$
case, with $n=6,7,8$, coincide~\rFLb\rWalg\ with the $c={6\o 7}, {7\o
10},{1\o 2}$ Virasoro minimal models, respectively, and therefore the
branching functions reduce to various Virasoro characters.

\vskip 2mm
Our results are as follows:
\vskip2mm

\no
${\bf A_n~(n\geq 1)}$: ~Here the quadratic form in \SofX\ reads
\eqn\qfAn{  {\bf m} C_{A_n}^{-1} {\bf m}^t ~=~
  {1\o n+1} \left( \sum_{a=1}^n a(n+1-a)m_a^2 + 2\sum_{1\leq a < b \leq n}
     a(n+1-b) m_a m_b \right)~~.}
For this algebra the generalization of \SofAiii\ is
\eqn\SQofAn{ S^Q_{A_n}(q) ~=~ q^{c_n/24} ~b^0_{2Q}(q;n) ~~,}
where the superscript $Q$ on $S$ indicates restriction of the
summation in \SofX\ to
\eqn\ResAn{\sum_{a=1}^n m_a a \equiv Q~({\rm mod}~n+1)}
with $Q=0,1,\ldots,n$.
The branching function on the rhs of \SQofAn\ corresponds to the
$\ZZ_{n+1}$-parafermi\-onic primary field of weight $Q(1-{Q\o n+1})$.
In general~\rDQ,
\eqn\blm{ \eqalign{b_m^l(q;n) ~=&~
  q^{l(l+2)/4(n+3)-m^2/4(n+1)+1/4(n+3)}\eta^{-2}   \cr
  &\times ~\sum_{r,s=0}^{\infty}
    {(-1)^{r+s}q^{r(r+1)/2+s(s+1)/2+r s(n+2)}} \cr
  &\times~ \Bigl\{q^{r(l+m)/2+s(l-m)/2}-q^{n+2-l+r[n+2-(l+m)/2]+s[n+2-(l-m)/2]}
    \Bigr\} \cr} }
when $l-m\in 2\ZZ$, $b_m^l=0$ otherwise, and
\eqn\Eta{\eta(q) ~=~ q^{1/24}\prod_{n=1}^\infty (1-q^n)}
 is the Dedekind eta function.
In the $n=1,2$ cases the branching  functions coincide with certain characters
$\chi^{(m)}_{r,s}(q)$
of the highest weight $\Delta^{(m)}_{r,s}={(r(m+1)-sm)^2-1\o 4m(m+1)}$
irreducible representations of the Virasoro algebra at central charge
$c=1-{6\o m(m+1)}$ ~\rRoCa
\eqn\Virchi{ \chi^{(m)}_{r,s}(q) ~=~ {1\o \eta(q)} \sum_{k\in \ZZ}
\left( q^{\Delta^{(m)}_{r+2km,s}} - q^{\Delta^{(m)}_{r+2km,-s}}\right)~~,}
namely,
\eqn\SofAlow{ \eqalign{ S_{A_1}^0(q) = \sum_{{m=0 \atop {\rm even}}}^\infty
  {q^{m^2/2} \o (q)_m} = q^{1/48} \chi_{1,1}^{(3)}(q) &,~~
  S_{A_1}^1(q) = \sum_{{m=1 \atop {\rm odd}}}^\infty
  {q^{m^2/2} \o (q)_m} = q^{1/48} \chi_{1,3}^{(3)}(q), \cr
  S_{A_2}^0(q) ~=~ q^{1/30}~[\chi^{(5)}_{1,1}(q)+\chi^{(5)}_{1,5}(q)]~&,
  ~~S_{A_2}^{\pm 1}(q) ~=~ q^{1/30}~\chi^{(5)}_{1,3}(q)~~.\cr} }

\vskip 2mm

When considered as characters of $(A_1^{(1)})_{n+1}/U(1)$, the
formula~\SQofAn\ is that of Lepowski and Primc~\rLepPrim.

\bigskip
\no
${\bf D_n~(n\geq 3)}$: ~In this case
\eqn\qfDn{ \eqalign{ {\bf m} C_{D_n}^{-1} {\bf m}^t ~=~
  & \sum_{a=1}^{n-2} a m_a^2 ~+~ {n\o 4}(m_{n-1}^2 ~+~ m_n^2)
  ~+~ 2\sum_{1\leq a < b \leq n-2} a m_a m_b \cr
  &+ ~ \sum_{a=1}^{n-2} a m_a(m_{n-1}+m_n)
  ~+~ {n-2\o 2} m_{n-1} m_n~~,\cr} }
and the result is
\eqn\SQofDn{ S^Q_{D_n}(q) ~=~ q^{1/24} ~f_{n,nQ}(q) }
with $Q=0,1$,
where summation is restricted to
\eqn\ResDn{m_{n-1}+m_n \equiv Q~({\rm mod}~2),}
 and the characters
\eqn\fdef{
   f_{n,j}(q) ~=~ {1\o \eta(q)} ~\sum_{k\in \ZZ} q^{n(k+{j\o 2n})^2}~~}
with $j=0,1,\ldots,n$ correspond to the extended primary fields of weight
${j^2 \o 4n}$.
Note that due to the coincidence $D_3=A_3$ the expressions \SQofAn\ and
\SQofDn\ are related when $n=3$ by (cf.~\rKedMc\rKaWa)
$S^0_{D_3}=S^0_{A_3}+S^2_{A_3}$ and $S^1_{D_3}=2S^1_{A_3}$.

\bigskip
\no
${\bf E_6}$: ~With a suitable labeling of roots we have
\eqn\carEviinv{ C_{E_6}^{-1} ~=~ \pmatrix{4/3 & 2/3& 1& 4/3& 5/3& 2\cr
                                         2/3 & 4/3& 1& 5/3& 4/3& 2\cr
                                         1&  1& 2& 2& 2& 3\cr
                                         4/3& 5/3& 2& 10/3& 8/3& 4\cr
                                         5/3& 4/3& 2& 8/3& 10/3& 4\cr
                                         2& 2& 3& 4& 4& 6\cr} ~.}
Here we find
\eqn\SQofEvi{ S^0_{E_6}(q) ~=~ q^{c/24}
  ~[\chi^{(6)}_{1,1}(q)+\chi^{(6)}_{5,1}(q)] ~~,
   ~~~~S^{\pm 1}_{E_6}(q) ~=~ q^{c/24}~ \chi^{(6)}_{3,1}(q) ~~, }
with the restrictions
\eqn\ResEvi{m_1-m_2+m_4-m_5 \equiv Q~({\rm mod}~3)}
 and $c=6/7$.
 The Virasoro characters are as in \Virchi,
the relevant weights being $\Delta_{1,1}^{(6)}=0$, $\Delta_{5,1}^{(6)}=5$,
and $\Delta_{3,1}^{(6)}=4/3$.

\bigskip
\no
${\bf E_7}$: ~Here we can write
\eqn\carEviiinv{ C_{E_7}^{-1} ~=~ \pmatrix{3/2& 1& 3/2& 2& 2& 5/2& 3 \cr
                                         1& 2& 2& 2& 3& 3& 4 \cr
                                         3/2& 2& 7/2& 3& 4& 9/2& 6 \cr
                                         2& 2& 3& 4& 4& 5& 6 \cr
                                         2& 3& 4& 4& 6& 6& 8 \cr
                                         5/2& 3& 9/2& 5& 6& 15/2& 9 \cr
                                         3& 4& 6& 6& 8& 9& 12 \cr} }
and
\eqn\SQofEvii{ S^0_{E_7}(q) ~=~ q^{c/24} ~\chi^{(4)}_{1,1}(q) ~~,
   ~~~~S^{1}_{E_7}(q) ~=~ q^{c/24} ~\chi^{(4)}_{3,1}(q) ~~. }
The restrictions in this case are
\eqn\ResEvii{m_1+m_3+m_6 \equiv Q~({\rm mod}~2),}
$c=7/10$, and $\Delta^{(4)}_{1,1}=0$, $\Delta^{(4)}_{3,1}=3/2$.

\bigskip \no
${\bf E_8}$: ~Finally in this case
\eqn\carEviiiinv{ C_{E_8}^{-1}~=~\pmatrix{2& 2& 3& 3& 4& 4& 5& 6 \cr
                                          2& 4& 4& 5& 6& 7& 8& 10 \cr
                                          3& 4& 6& 6& 8& 8& 10& 12 \cr
                                          3& 5& 6& 8& 9& 10& 12& 15 \cr
                                          4& 6& 8& 9& 12& 12& 15& 18 \cr
                                          4& 7& 8& 10& 12& 14& 16& 20 \cr
                                          5& 8& 10& 12& 15& 16& 20& 24 \cr
                                          6& 10& 12& 15& 18& 20& 24& 30 \cr} }
and
\eqn\SofEviii{ S_{E_8}(q) ~=~ q^{c/24} ~\chi^{(3)}_{1,1}(q) ~~,}
is the character of the identity ($\Delta=0$) representation of the
Virasoro algebra with $c=1/2$.

\bigskip
The above generalizations of~\rKedMc\ have been verified using
Mathematica$^{{\rm TM}}$ for
all simply-laced Lie algebras of rank $n\leq 8$ up to order $q^{100}$.

It would of course be most desirable to have direct proofs of the
$D_n$ and $E_n$ sum representations for the characters, as for the $A_n$
case~\rLepPrim.

\bigskip
\bigskip

\newsec{Quasi-Particle Representation}

 All the above representations  for branching functions may by
put in a fermionic quasi-particle form by following the reverse of the
procedure of ~\rKedMc. Let $P_d(m,N)$ denote the
number of
ways that a positive integer $N$ can be additively
partitioned into $m$ distinct positive integers. (We set
$P_d(0,0)=1$ and $P_d(m,0)=P_d(0,N)=0$ ~for $m,N>0$.) Then using
\eqn\genPd{ \sum_{N=0}^\infty ~P_d(m,N) ~q^N ~=~ {q^{m(m+1)/2} \o (q)_m} }
we rewrite \SofX\ as
\eqn\SofXa{ S_G(q) = \sum_{{\bf N=0}}^\infty
                             ~\sum_{{\bf m=0}}^\infty
  \left( \prod_{a=1}^n P_d(m_a,N_a) \right)
    ~q^{\sum_{a=1}^n [ N_a + m_a
 ( \sum_{b=1}^n (C_G^{-1})_{ab}m_b -{m_a+1\o 2} ) ] } ~. }

Fermionic quasi-particles are characterized by a set of single
particle energy levels $e_a(P)$ and an additive composition law
for multi-particle energy gaps above the ground state
\eqn\En{\hat{E}(\{P\}) ~\equiv~
     E_{{\rm ex}}(\{P\})-E_{GS}~=\sum_{a,j_a,{\rm rules}}e_a(P_{a,j_a}) ~~,}
and their momenta
\eqn\Mom{P(\{P\})-P_{GS}~=\sum_{a,j_a,{\rm rules}}P_{a,j_a} ~.}
One of the rules of composition is the fermionic exclusion rule
\eqn\FerEx{P_{a,i_a}\neq P_{a,j_a}\qquad\hbox{for}\quad i_a\neq j_a~.}
To specify the $P_{a,j_a}$ we consider the system to be in a box of size
$L$ which serves as an infrared cutoff and quantizes the
 $P_{i}$ in units of $2\pi/L$.
Taking a continuum limit, the massless
single particle energies $e_a$ all reduce to
\eqn\disp{e_{a}(P) ~=~ v|P| ~~,}
where $v$ is called the speed of sound (or light). Considering
the partition function for left moving excitations, with the ground
state energy scaled out,
 \eqn\Part{Z(q)~=~\sum e^{-\hat{E}/kT},}
and setting
\eqn\qDef{q~=~e^{-2\pi v/L kT} ~,}
we see that \SofXa\ results if the $P_{a,j_a}$ are restricted
by
\eqn\Momdef{P_{a,j_a}~=~P^{{\rm min}}_a({\bf m})+{2\pi\o L}k_{j_a}}
where $k_{j_a}$ are distinct nonnegative integers and
\eqn\Pmin{  P^{{\rm min}}_a({\bf m})~ =~ {2\pi\o L}
\Biggr[{1\o 2}
   ~+~ {1\o 2} \sum_{b=1}^n N_{ab}  m_b\Biggl], }
in terms of the matrix $N=2 C^{-1}_G - 1$.
This generalizes the infrared anomaly rule (3.15) of~\rKedMc.

\vskip 4mm

A particularly interesting example of the quasi-particle
representation is provided by comparing the $E_8$ sum with the $Q=0$
sector of the $A_1$ sum, both of which give the identity character of
the critical Ising model. This demonstrates the important fact that
the fermionic quasi-particle representation is not unique, in general.
In the $A_1$ case ~$C_{A_1}=2$, so the momentum restrictions~\Pmin\
are simply those of a free fermion, $P^{{\rm min}}= {\pi/L}$
independent of ${\bf m}$.  In contrast, the $E_8$ representation
involves 8 quasi-particles with nontrivial momentum restrictions.  We
compare these representations in tables~1 and~2, where we show the
particle content up to order $q^{30}$.  The order $q^{30}$ is chosen
because this is the lowest order where all $E_8$ quasi-particle types
make a nonvanishing contribution.  One pattern that is seen to emerge
from the two tables, namely that the lowest level where a
$2N$-particle state appears in the $A_1$ spectrum is also the lowest
level a state of $N$ quasi-particles of type 1 appears in the $E_8$
spectrum, is easily shown to persist for all $N$.  In table~3 we give
more details about the structure of the $E_8$ representation up to
order $q^{14}$.

\bigskip

The meaning of the restrictions~\ResAn, \ResDn, \ResEvi, \ResEvii\ is
understood in the above picture if we assign the quasi-particles
charges with respect to a symmetry of the Dynkin diagram of the affine
extension $G^{(1)}$ of $G$.  In the $A_n$ case this symmetry is $\ZZ_2
\semidir \ZZ_{n+1}$ (except for the $n$=1 case where it is just
$\ZZ_2$) and the quasi-particle of type $a$ carries a $\ZZ_{n+1}$
charge $a$, quasi-particles $a$ and $n+1-a$ being $C$-conjugates of
each other. Thus $Q=\sum_{a=1}^n m_a a$ is the total charge of the
system, and $S^Q_G(q)$ is the partition function in the corresponding
charge sector. In the $D_n$ case we need to utilize just a $\ZZ_2$
subgroup of the full symmetry of the $D_n^{(1)}$ Dynkin diagram, under
which the quasi-particles $n-1$ and $n$ are charged while all others
are neutral. In the $E_6$ case the symmetry is $\ZZ_2
\semidir \ZZ_3$, with the $\ZZ_3$-charge assignment $1,-1,0,1,-1,0$
for the quasi-particles $a=1,\ldots,6$, respectively. Finally, the
$E_7$ system exhibits a $\ZZ_2$ symmetry, under which quasi-particles
1, 3, and 6 are odd, the rest being even.  It can easily be seen that
${\bf m} C_G^{-1} {\bf m}^t ({\rm mod}~1)$ is constant in every charge
sector --- as it should, since the branching functions have a power
series expansion in $q$ (up to an overall possibly fractional power)
--- while generically they are different in sectors of different
charge.

\bigskip

\newsec{Extracting the Central Charge}
The leading behavior of any character $\chi(q)$ of a
CFT for $q\to 1^-$ is determined~\rKedMc\rDV\rNRT\ by the
effective central charge $\tilde{c}=c-12d_{{\rm min}}$
(where $d_{{\rm min}}$ is the smallest scaling dimension in the theory):
\eqn\SGlim{\chi(q) ~\simeq ~ \tilde{q}^{-\tilde{c}/24}
       ~~~~~~~~{\rm as}~~q\to 1^- ,}
where $\tilde{q}=e^{-2\pi L kT/v}$ for $q$ given by~\qDef.
In the cases at hand, where the characters are of the form~\SofX,
$\tilde{c}$ can be evaluated
along the lines of~\rRiSz. The result can be written as follows
(cf.~\rNRT): For an $n$ by $n$ symmetric matrix $B$,
\eqn\SofB{ \sum_{{\bf m=0}}^{\infty} ~
    {q^{{\bf m} B {\bf m}^t} \o (q)_{m_1}\ldots(q)_{m_n} } ~\simeq~
   \tilde{q}^{-\tilde{c}/24}\qquad{\rm as}~~q\to 1^- ~,}
where
\eqn\ctilde{ \tilde{c} ~=~ {6\o \pi^2} \sum_{a=1}^n
  {\cal L}\left({x_a \o 1+x_a}\right)~~.}
Here
\eqn\dilog{ {\cal L}(z) ~=~ -{1\o 2} \int_0^z dt \left[
   {\ln t \o 1-t} + {\ln (1-t) \o t} \right]   }
is the Rogers dilogarithm function~\rLewin, and the $x_a$ satisfy
the equations
\eqn\xeq{  x_a ~=~ \prod_{b=1}^n (1+x_b)^{-N_{ab}}~~, }
with $N=2B-1$.

In the cases of interest here, where $B$ is the inverse Cartan
matrix of a simply-laced algebra $G$,
eqs.~\ctilde\ and~\xeq\ have been
encountered and solved previously in the context of
1) purely elastic scattering theories and their ultraviolet limits~\rourii
{}~and 2) the finite-temperature thermodynamics of RSOS spin chains~\rKBR .
Eqs.~\xeq\ have a unique real solution. Inserting this solution
into~\ctilde\ and using appropriate sum rules~\rsums\rKBR\rourii
{}~for the Rogers dilogarithm
the known
central charges
of the ${(G^{(1)})_1\times(G^{(1)})_1 \o (G^{(1)})_2}$ coset CFTs
are obtained.

\vskip 3mm

In~\rNRT\ eqs.~\SGlim--\xeq\ were applied to the non-unitary Virasoro
minimal models $M(2,2n+3)$ of central charge $c=-{2n(6n+5)\o 2n+3}$.
The sum side of the Andrews-Gordon identities~\rGord\  takes the
role of $S_G(q)$ in this case,
the product side being a way of writing the characters of the primary
fields in these CFTs~\rFNO.
The sum for the character of the smallest dimension field is
 of the form~\SofX\ if one
takes $(2-I_n)^{-1}$ instead of the inverse Cartan matrix,
where $I_n$ is the ``generalized incidence matrix''
for the twisted affine Lie algebra $A_{2n}^{(2)}$ introduced in~\rourii\
(it differs from that of the Lie algebra $A_n$ only by a 1 on the last
entry of the diagonal, \ie~it is the incidence matrix of the tadpole
graph $A_{2n}/{\bf Z}_2$).
As a generalization of an observation made in~\rKedMc ,
we note that this sum can be obtained from the one in $S_{D_{n+2}}(q)$
by avoiding the summation over $m_{n+1}$ and $m_{n+2}$ (\ie~setting
both to zero) due to the fact that the upper-left $n\times n$ minor
of $C_{D_{n+2}}^{-1}$ is precisely $(2-I_n)^{-1}$.
Eqs.~\SGlim--\xeq\ then give the effective
central charges $\tilde{c}={2n\o 2n+3}$
of these
theories, cf.~\rourii\rNRT.
Our discussion in sect.~3 extends in a straightforward way to give
a quasi-particle interpretation of the models $M(2,2n+3)$ (cf.~\rKedMc\
for the case $n$=1).

\bigskip

\newsec{Discussion}

In~\rKedMc\ the fermionic quasi-particle representation~\SofAiii\ of the
branching functions was found by studying the spectrum of the
antiferromagnetic 3-state Potts chain, realized as
an RSOS model based on the Dynkin diagram of
$D_{4}$. This $D_4$ model is obtained by
an orbifold construction~\rFG\rPZ\ from the $A_{5}$ RSOS
model, and the antiferromagnetic spectrum corresponds to the boundary
of the I/II regime. It is expected that the
fermionic representation of the branching functions given above for
$A_{n}$ can be obtained from a similar study of the $A_{n+2}$ RSOS
models at the boundary of the I/II regime. Such a study would extend
the computations of Kl\"umper and Pearce~\rKlumPear\  and would provide
explicit results for the other characters not considered in this
paper. We remark, however, that the analagous study of the
ferromagnetic region of the $\ZZ_{n+1}$
model of Fateev and
Zamolodchikov~\rFatZam,  which has the identical $\ZZ_{n+1}$ parafermionic
branching functions as the I/II boundary of the $A_{n+2}$ RSOS
models, gives a different fermionic representation~\rDasKedMc\  because
the quasi-particle spectra of the lattice models are quite different.

The $ADE$ algebras that appear in the fermionic sum formulas presented
here are the same as the  algebras relevant for the integrable
massive scattering theories discussed in~\rourii, which describe
certain perturbations of the corresponding $ADE$-related coset CFTs.
For example, the energy perturbation of the critical Ising model, the
well known Ising field theory~\rMTW, is related to $A_1$, and the
magnetic perturbation of the Ising model describes a scattering theory
with 8 particles, as discovered by A.~Zamolodchikov~\rTanig.  Not only
do the numbers of particles in the massive scattering theories agree
with those in the quasi-particle picture, also the charge assignments
for the quasi-particles in sect.~3 are consistent with those of the
particles in the scattering theories.
Furthermore, we note
the similarity of~\Momdef, \Pmin\ with the UV limit (large rapidities) of
the Bethe Ansatz equations for the diagonal scattering theories
discussed in~\rourii.

We
therefore
strongly suspect that different fermionic representations of the
same branching functions are closely related to various integrable
perturbations of the relevant CFT.    It is also interesting to see
whether there is any relation between the massless quasi-particle
interpretations suggested by our results and the massless $S$-matrix
theories which were associated with certain CFTs in~\rZams\
(where different $S$-matrices for a given CFT were also thought of as
corresponding to different integrable perturbations of that CFT).
We believe that characters of
${(G^{(1)})_1 \times (G^{(1)})_1 \o (G^{(1)})_2}$ not discussed in
this paper also have fermionic representations, as in the $A_n$ case
discussed in~\rLepPrim.

\vskip 20mm

{{\vbox{\centerline{\bf Acknowledgements}}}
\nobreak\medskip
We are pleased to acknowledge fruitful discussions with Prof. J.
Lepowsky.
The work of R.K.~and B.M.M.~was partially supported by the
National Science Foundation under grant DMR-9106648.
The work of T.R.K.~was supported by NSERC and the Department of Energy,
grant DE-FG05-90ER40559, and
that of E.M.~by NSF grant 91-08054.
E.M.~wishes to thank the High Energy group at Tel Aviv University
for their hospitality during the time in which part of this work has
been done.}

\bigskip

\vfill
\eject
\listrefs


\setbox\strutbox=\hbox{\vrule height9pt depth5pt width0pt}
\centerline{\vbox{\halign{&#\vrule&\strut~#&
                           #\vrule&\strut~#&
                           #\vrule&\strut~#&
                           #\vrule&\strut~#&
                           #\vrule&\strut~#&
                           #\vrule\cr\tablerule
& ~~$\hat{E}~L/2\pi v$~~ && ~~ $d_{\hat{E}}$ ~~
&& 2-{\rm particle}  && 4-{\rm particle}
&& 6-{\rm particle} & \cr \tablerule\tablerule
& 0   && 1   &&     &&     &&  {}   &\cr\tablerule
& 1   && 0   &&     &&     &&  {}   &\cr\tablerule
& 2   && 1   &&  1  &&     &&  {}   &\cr\tablerule
& 3   && 1   &&  1  &&     &&  {}   &\cr\tablerule
& 4   && 2   &&  2  &&     &&  {}   &\cr\tablerule
& 5   && 2   &&  2  &&     &&  {}   &\cr\tablerule
& 6   && 3   &&  3  &&     &&  {}   &\cr\tablerule
& 7   && 3   &&  3  &&     &&  {}   &\cr\tablerule
& 8   && 5   &&  4  &&  1  &&  {}   &\cr\tablerule
& 9   && 5   &&  4  &&  1  &&  {}   &\cr\tablerule
& 10   && 7   && 5   && 2   && {}   &\cr\tablerule
& 11   && 8   && 5   && 3   && {}   &\cr\tablerule
& 12   && 11  && 6   && 5   && {}   &\cr\tablerule
& 13   && 12  && 6   && 6   && {}   &\cr\tablerule
& 14   && 16  && 7   && 9   && {}   &\cr\tablerule
& 15   && 18  && 7   && 11  && {}   &\cr\tablerule
& 16   && 23  && 8   && 15  && {}   &\cr\tablerule
& 17   && 26  && 8   && 18  && {}   &\cr\tablerule
& 18   && 33  && 9   && 23  &&  1   &\cr\tablerule
& 19   && 37  && 9   && 27  &&  1   &\cr\tablerule
& 20   && 46  && 10  && 34  &&  2  &\cr\tablerule
& 21   && 52  && 10  && 39  &&  3  &\cr\tablerule
& 22   && 63  && 11  && 47  &&  5  &\cr\tablerule
& 23   && 72  && 11  && 54  &&  7  &\cr\tablerule
& 24   && 87  && 12  && 64  &&  11  &\cr\tablerule
& 25   && 98  && 12  && 72  &&  14  &\cr\tablerule
& 26   && 117 && 13  && 84  &&  20  &\cr\tablerule
& 27   && 133 && 13  && 94  &&  26  &\cr\tablerule
& 28   && 157 && 14  && 108 &&  35  &\cr\tablerule
& 29   && 178 && 14  && 120 &&  44  &\cr\tablerule
& 30   && 209 && 15  && 136 &&  58  &\cr\tablerule}}}
\vskip 1mm
\no {{\bf Table~1:}
Total, $d_{\hat{E}}$, and number of 2-, 4-, and 6-particle states
contributing to the $A_1$ representation of the character
$q^{1/48} \chi_{1,1}^{(3)}$ of the critical Ising model.}

\vfill\eject

\setbox\strutbox=\hbox{\vrule height9pt depth5pt width0pt}
\centerline{\vbox{\halign{&#\vrule&\strut~#&
               #\vrule&\strut~#&
               #\vrule&\strut~#&
               #\vrule&\strut~#&
               #\vrule&\strut~#&
               #\vrule\cr\tablerule
& ~~$\hat{E}~L/2\pi v$~~ && ~ $d_{\hat{E}}$ ~
&& 1-{\rm particle}  && 2-{\rm particle} && 3-{\rm particle}
&\cr\tablerule\tablerule
& 0   && 1   &&     &&     &&  {}   &\cr\tablerule
& 1   && 0   &&     &&     &&  {}   &\cr\tablerule
& 2   && 1   &&  1  &&     &&  {}   &\cr\tablerule
& 3   && 1   &&  1  &&     &&  {}   &\cr\tablerule
& 4   && 2   &&  2  &&     &&  {}   &\cr\tablerule
& 5   && 2   &&  2  &&     &&  {}   &\cr\tablerule
& 6   && 3   &&  3  &&     &&  {}   &\cr\tablerule
& 7   && 3   &&  3  &&     &&  {}   &\cr\tablerule
& 8   && 5   &&  4  &&  1  &&  {}   &\cr\tablerule
& 9   && 5   &&  4  &&  1  &&  {}   &\cr\tablerule
& 10   && 7   && 4   && 3   && {}   &\cr\tablerule
& 11   && 8   && 4   && 4   && {}   &\cr\tablerule
& 12   && 11  && 5   && 6   && {}   &\cr\tablerule
& 13   && 12  && 5   && 7   && {}   &\cr\tablerule
& 14   && 16  && 6   && 10   && {}   &\cr\tablerule
& 15   && 18  && 6   && 12  && {}   &\cr\tablerule
& 16   && 23  && 6   && 17  && {}   &\cr\tablerule
& 17   && 26  && 6   && 20  && {}   &\cr\tablerule
& 18   && 33  && 6   && 26  &&  1   &\cr\tablerule
& 19   && 37  && 6   && 30  &&  1   &\cr\tablerule
& 20   && 46  && 7   && 36  &&  3  &\cr\tablerule
& 21   && 52  && 7   && 40  &&  5  &\cr\tablerule
& 22   && 63  && 7   && 48  &&  8  &\cr\tablerule
& 23   && 72  && 7   && 54  &&  11  &\cr\tablerule
& 24   && 87  && 7   && 64  &&  16  &\cr\tablerule
& 25   && 98  && 7   && 71  &&  20  &\cr\tablerule
& 26   && 117 && 7   && 82  &&  28  &\cr\tablerule
& 27   && 133 && 7   && 90  &&  36  &\cr\tablerule
& 28   && 157 && 7   && 102 &&  48  &\cr\tablerule
& 29   && 178 && 7   && 111 &&  60  &\cr\tablerule
& 30   && 209 && 8   && 123 &&  78  &\cr\tablerule}}}
\vskip 2mm
\no {{\bf Table~2:}
Total, $d_{\hat{E}}$, and number of 1-, 2-, and 3-particle states
contributing to the $E_8$ representation of the character
$q^{1/48} \chi_{1,1}^{(3)}$ of the critical Ising model.}

\vfill\eject

\setbox\strutbox=\hbox{\vrule height8.2pt depth5pt width0pt}
\centerline{\vbox{\halign{&#\vrule&\strut$~#~$&
               #\vrule&\strut$~#~$&
               #\vrule&\strut$~#~$&
               #\vrule&\strut$~#~$&
               #\vrule&\strut$~#~$&
               #\vrule\cr\tablerule
& ~\hat{E}~L/2\pi v~ && ~~ d_{\hat{E}} ~~
&& ~~~~~~~~~~~~~{\bf m}~~~~~~~~~~~~~ && ~ d_{{\hat{E},{\bf m}}} ~
&& ~~~~~~~{\bf P}^{\rm min}({\bf m})~L/2\pi ~~~~~~~
&\cr\tr\tr
& 2   && 1   && (1,0,0,0,0,0,0,0) && 1 && (2,0,0,0,0,0,0,0) &\cr\tr\tr
& 3   && 1   && (1,0,0,0,0,0,0,0) && 1 && (2,0,0,0,0,0,0,0) &\cr\tr\tr
& 4   && 2   && (1,0,0,0,0,0,0,0) && 1 && (2,0,0,0,0,0,0,0) &\cr\tr
&     &&     && (0,1,0,0,0,0,0,0) && 1 && (0,4,0,0,0,0,0,0) &\cr\tr\tr
& 5   && 2   &&       ....        &&   &&       ....        &\cr\tr\tr
& 6   && 3   &&       ....        &&   &&       ....        &\cr\tr
&     &&     && (0,0,1,0,0,0,0,0) && 1 && (0,0,6,0,0,0,0,0) &\cr\tr\tr
& 7   && 3   &&       ....        &&   &&       ....        &\cr\tr\tr
& 8   && 5   &&       ....        &&   &&       ....        &\cr\tr
&     &&     && (0,0,0,1,0,0,0,0) && 1 && (0,0,0,8,0,0,0,0) &\cr\tr
&     &&     && (2,0,0,0,0,0,0,0) && 1 && ({7\o 2},0,0,0,0,0,0,0) &\cr\tr\tr
& 9   && 5   &&       ....        &&   &&       ....        &\cr\tr\tr
& 10  && 7   &&       ....        &&   &&       ....        &\cr\tr
&     &&     && (2,0,0,0,0,0,0,0) && 2 && ({7\o 2},0,0,0,0,0,0,0) &\cr\tr
&     &&     && (1,1,0,0,0,0,0,0) && 1 && (4,6,0,0,0,0,0,0) &\cr\tr\tr
& 11  && 8   &&       ....        &&   &&       ....        &\cr\tr
&     &&     && (2,0,0,0,0,0,0,0) && 2 && ({7\o 2},0,0,0,0,0,0,0) &\cr\tr
&     &&     && (1,1,0,0,0,0,0,0) && 2 && (4,6,0,0,0,0,0,0) &\cr\tr\tr
& 12  && 11  &&       ....        &&   &&       ....        &\cr\tr
&     &&     && (2,0,0,0,0,0,0,0) && 3 && ({7\o 2},0,0,0,0,0,0,0) &\cr\tr
&     &&     && (1,1,0,0,0,0,0,0) && 3 && (4,6,0,0,0,0,0,0) &\cr\tr
&     &&     && (0,0,0,0,1,0,0,0) && 1 && (0,0,0,0,12,0,0,0) &\cr\tr\tr
& 13  && 12  &&       ....        &&   &&       ....        &\cr\tr
&     &&     && (2,0,0,0,0,0,0,0) && 3 && ({7\o 2},0,0,0,0,0,0,0) &\cr\tr
&     &&     && (1,1,0,0,0,0,0,0) && 4 && (4,6,0,0,0,0,0,0) &\cr\tr\tr
& 14  && 16  &&       ....        &&   &&       ....        &\cr\tr
&     &&     && (2,0,0,0,0,0,0,0) && 4 && ({7\o 2},0,0,0,0,0,0,0) &\cr\tr
&     &&     && (1,1,0,0,0,0,0,0) && 5 && (4,6,0,0,0,0,0,0) &\cr\tr
&     &&     && (0,0,0,0,0,1,0,0) && 1 && (0,0,0,0,0,14,0,0) &\cr\tr
&     &&     && (1,0,1,0,0,0,0,0) && 1 && (5,0,9,0,0,0,0,0) &\cr\tr\tr}}}
\vskip 0.4mm
\no {{\bf Table~3:} More details about the structure of the $E_8$
representation of table~2, namely the occupation number vectors
${\bf m}$ contributing for a given energy gap
$\hat{E}$, the degeneracy $d_{\hat{E},{\bf m}}$
at given $\hat{E}, {\bf m}$, and the
minimal momenta $P^{\rm min}_a({\bf m})$ assembled into a vector
${\bf P}^{\rm min}({\bf m})$, with $P^{\rm min}_a({\bf m})\equiv 0$ if
$m_a=0$. For $\hat{E}>4$ we only show ${\bf m}$ with $d_{\hat{E},{\bf m}}>1$
or those that have not appeared for smaller $\hat{E}$.}

\vfill\eject

\bye\end